\documentclass{article}

\usepackage{arxiv}

\usepackage[utf8]{inputenc} 
\usepackage[T1]{fontenc}    
\usepackage{hyperref}       
\usepackage{url}            
\usepackage{booktabs}       
\usepackage{amsfonts}       
\usepackage{nicefrac}       
\usepackage{microtype}      
\usepackage{graphicx}
\usepackage{amsthm,amsmath,amssymb}
\usepackage{subcaption,tikz}
\usepackage[authoryear]{natbib}

\usetikzlibrary{fit,positioning}

\title{Space-Time VON CRAMM: Evaluating Decision-Making in Tennis with Variational generatiON of Complete Resolution Arcs via Mixture Modeling}

\author{Stephanie Kovalchik\thanks{Author website \texttt{on-the-t.com}} \\
  Zelus Analytics\\
  Austin, Texas USA \\
  \texttt{skovalchik@zelusanalytics.com} \\
  \And
  Martin Ingram \\
  University of Melbourne \\
  Melbourne, Australia \\
  \texttt{martin.ingram@gmail.com} \\
  \And
  Kokum Weeratunga\\
  Edge 10 \\
  Melbourne, Australia \\
  \texttt{kokum.weeratunga@gmail.com} \\
    \And
  Cagatay Goncu \\
  Monash University \\
  Melbourne, Australia \\
  \texttt{Cagatay.Goncu@monash.edu} 
}

\date{}

\renewcommand{\headeright}{Technical Report}
\renewcommand{\undertitle}{Technical Report}

\hypersetup{
pdftitle={},
pdfsubject={stat.ME},
pdfauthor={Stephanie Kovalchik, Martin Ingram, Kokum Weeratunga, Cagatay Goncu},
pdfkeywords={First keyword, Second keyword, More},
}

\begin{document}
\maketitle

\begin{abstract}
  Sports tracking data are the high-resolution spatiotemporal observations of a competitive event. The growing collection of these data in professional sport allows us to address a fundamental problem of modern sport: how to attribute value to individual actions? Taking advantage of the smoothness of ball and player movement in tennis, we present a functional data framework for estimating expected shot value (ESV) in continuous time. Our approach is a three-step recipe: 1) a generative model for a full-resolution functional representation of ball and player trajectories using an infinite Bayesian Gaussian mixture model (GMM), 2) conditioning of the GMM on observed positional data, and 3) the prediction of shot outcomes given the functional encoding of a shot event. 
From the ESV we derive three metrics of central interest:  value added with shot taking (VAST), Shot IQ, and value added with court coverage (VACC), which respectively attribute value to shot execution, shot selection and movement around the court. We rate player performance at the 2019 US Open on these advanced metrics and show how each adds a novel perspective to performance evaluation in tennis that goes beyond simple counts of outcomes by quantitatively assessing the decisions players make throughout a point.
\end{abstract}


\section{Introduction}

One of the biggest stories to come out of the 2019 US Open Tennis Championships, the last of the four biggest events of the professional tennis calendar known as the Grand Slams, was Serena Williams' loss in the finals to 19 year-old Bianca Andreescu. The debate that this surprise loss created put a spotlight on one of the biggest challenges for statistical analysis in sports: the problem of attribution. Given the myriad of events that happen from the time play starts until a change in score occurs, how can we \emph{attribute} a score to any one or more of those actions? Such attribution requires going beyond box scores; it requires the quantification  of the micro-decisions that happen over the space and time evolution of a competitive event.

The advent of modern tracking data  has made fine-grained spatiotemporal analysis a reality in sport.  Although tracking data has been a mainstay of professional tennis matches over the past decade, analysis with these data has rarely gone beyond simple descriptions of ball landing and impact locations. A few exceptions are several prediction models that have been proposed for relating shot and player features to a shot's outcome \citep{wei2013sweet,wei2016thin} or bounce location \citep{wei2013predicting, wei2016forecasting}. A recent study used a generative adversarial network to generate descriptions of a shot and player positions in the form of 2D flattened images \citep{fernando2019memory}.

A few exceptions are several prediction models that have been proposed for relating shot and player features to a shot’s outcome \citep{wei2013sweet,wei2016thin}  or bounce location \citep{wei2013predicting, wei2016forecasting}. Another work assigned shot value during a point based on a Markovian model whose state transitions are based on coarsened locations of the ball and players at one time point \citep{floydshot}. And a recent study used a generative adversarial network to generate descriptions of a shot and player positions in the form of 2D flattened images \citep{fernando2019memory}.

These recent developments have been important in producing more advanced metrics of tennis performance. But, when it comes to the challenges of attribution, they don’t go far enough because none of these models account for all of the events that happen throughout the space and time of a point. It is only with spatiotemporal models of the continuous sequence of events during a point that it becomes possible to decouple the separate contributions those events have on score outcomes.

\begin{figure}
     \centering
     \begin{subfigure}[b]{\textwidth}
         \centering
         \includegraphics[width=0.8\textwidth]{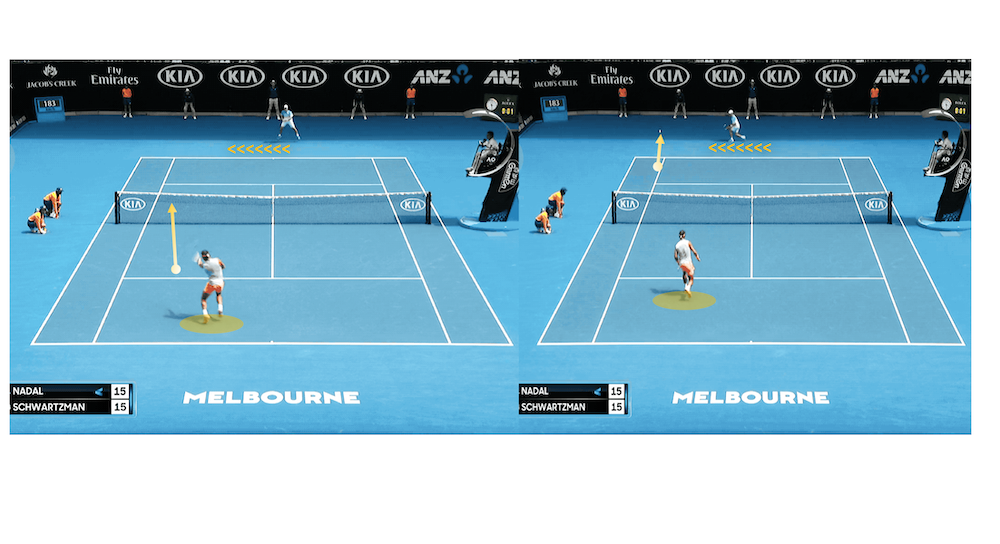}
         \caption{Rafael Nadal forehand down-the-line against Diego Schwartzman in the Round of 16 at the 2018 Australian Open. The ball eventually landed out, losing the point for Nadal.}
         \label{fig:point1}
     \end{subfigure}
     \hfill
     \begin{subfigure}[b]{\textwidth}
         \centering
         \includegraphics[width=0.8\textwidth]{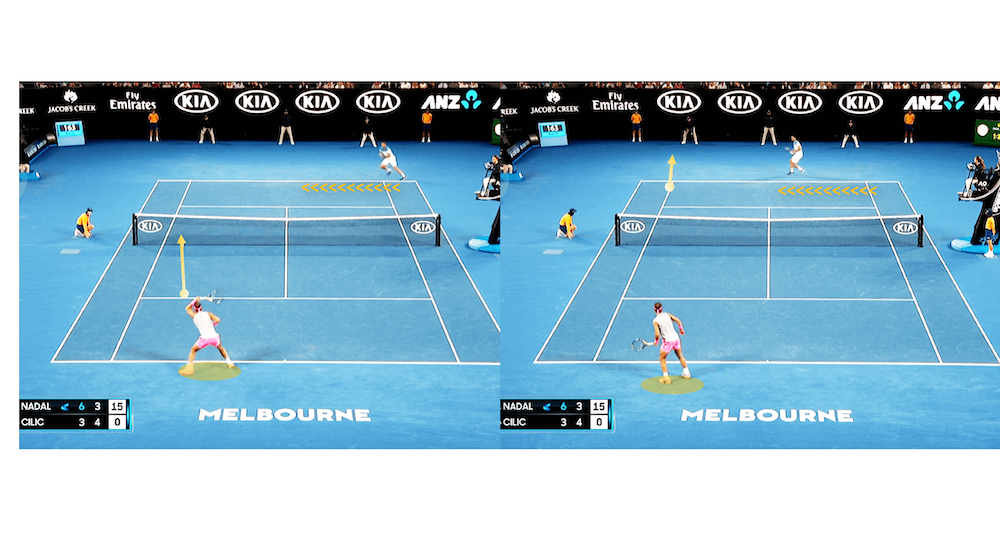}
         \caption{Rafael Nadal forehand down-the-line against Marin Cilic in the quarterfinal of the 2018 Australian Open. The shot was a clean winner in this case.}
         \label{fig:point2}
     \end{subfigure}
        \caption{Similar serve + 1 shots hit by Nadal illustrating how player position and court coverage can impact the outcome of otherwise identical shots.}
        \label{fig:shot_example}
\end{figure}

To make this more concrete consider the two shot scenarios depicted in Figure~\ref{fig:shot_example}. Both the top and bottom images show the start (on the left) and end (on the right) of nearly identical down-the-line forehands hit by Rafael Nadal at the 2018 Australian Open. What is most distinctive about the scenarios is the behavior of the receiving player. On the top panel, Diego Schwartzman is the opponent, a player considered to be one of the strongest defenders in the sport. Schwartzman is in a central position, several meters behind the baseline and beginning to move to the right when Nadal makes the shot. On the bottom panel, the receiver is Marin Cilic, who is far out wide and only inches behind the baseline when Nadal makes the shot. Nadal's shot goes out against Schwartzman but lands just inside the line against Cilic. The question is whether Schwartzman's positioning put more pressure on Nadal and increased the chance of his shot going out? Answering that question requires a generative model of receiver events, conditional on the shot they received, and the ability to predict a shot's outcome given each possible response by the receiver. No previous work in tennis has developed a tool with these two fundamental properties.

While tennis has lagged behind, other sports are experiencing a boom in spatiotemporal modeling. In the past 5 years, some of the most interesting examples have come out of approaches to measure expected possession value (EPV) in team sports. EPV is distinctive among most stats used in sport in providing a measure of performance that continuously updates with each action of a possession. The complexity of the actions and their configuration across multiple players on the court makes it a challenge to build probability models that can describe them and their association with score outcomes. \citet{cervone2016multiresolution} developed a multiresolution approach that consists of a low-resolution Markov chain whose transitions are driven by high-resolution microtransitions that link detailed space and time features to the major events of a possession. Recently, \citet{fernandez2019decomposing} presented an EPV approach for soccer that uses several deep learning models of the major actions in soccer and combines them into a cohesive stochastic process describing the  spatiotemporal evolution of the possession.

Inspired by these EPV models for team sports, the present paper presents a model for estimating expected shot value (ESV) in tennis. A key point of distinction with our approach compared to past EPV methods is the encoding of shot and player trajectories into a lower-dimensional functional representation that retains all available space-time information. We show that an infinite Bayesian Gaussian mixture model provides a generative distribution for the 3D ball and 2D player trajectories that is in good agreement with real data. A practical strength of this approach is that it can be fit using highly scalable variational inference methods, which we exploit to train models on 125,000 men’s shots and 80,000 women’s shots sampled from past Australian Opens. We derive conditional distributions for the generative model in order to project all possible future paths for the ball and players from any time point into a ball’s flight. The estimation of shot value is completed by linking each future to an outcome classification (‘win’, ‘error’, or ‘in play’) that is predicted from hierarchical generalized additive models with non-linear spatial effects. Gottfried Von Cramm, a legend of early 20th century tennis known for his `picture perfect' game\footnote{\url{https://www.tennisfame.com/hall-of-famers/inductees/baron-gottfried-von-cramm}}, provides the backronym that encapsulates the major features of our framework: Variational generatiON of Complete Resolution Arcs via Mixture Modeling (VON CRAMM).

\section{Expected Shot Value}

We will use $\mathcal{A}(\omega)$ to denote the full-resolution functional encoding of the shot event $\omega$. Here a `shot event' will refer to all of the actions of the ball and players during a shot. To be a full-resolution encoding, this lower-dimensional representation of the shot must retain all of the spatial-temporal information contained in $\omega$. 

Let $\mathbf{X}_t$ be the high-dimensional vector of the observed spatial-temporal information about a shot event up to time $t$ into the shot and let $W(\omega) = \lbrace 0, 1 \rbrace$ be an indicator that the shot wins the point. Our goal is to use this information to estimate the ESV at any time $t$, 

\begin{equation}
ESV_t = E[W(\omega) | \mathbf{X}_t],\; t \geq 0
\label{eq:epv}
\end{equation}

\noindent Given that the stochastic encoding $\mathcal{A}(\omega)$ contains all of the information about the duration of the shot, and the paths of the ball and player trajectories during the shot; the $ESV$ can be represented in terms of the conditional relationship between $\mathcal{A}(\omega)$ given the observed data $\mathbf{X}_t$,

\begin{equation}
ESV_t = \int_{\psi \in \Omega} \mathbb{P}{(W(\omega) | \mathcal{A}(\omega) = \mathcal{A}(\psi) )} \mathbb{P}(\mathcal{A}(\omega) =  \mathcal{A}(\psi) |\mathbf{X}_t)  d\psi
\label{eq:full}
\end{equation}

\noindent where $\mathbb{P}(.)$ denotes a probability density and $\Omega$ is the space of all possible shot events. 

Equation~(\ref{eq:full}) shows us that calculating the full-resolution $ESV$ at any time during a shot requires a generative distribution for $\mathcal{A}(\omega)$, the conditional distribution for $\mathcal{A}(\omega) | \mathbf{X}_t$, and predicted outcomes for $W(\omega)$. Thus, ESV can be likened to a recipe with three ingredients: 1) a generative model of the space-time dynamics of a shot event, 2) conditional generation of the shot event given observed data, 3) and outcome prediction given the specific features of a shot event. The goal of the present work is to develop a strategy for each of these ingredients.

\section{The VON CRAMM Recipe}

\subsection{Generative Model}

To obtain $\mathcal{A}(\omega)$ we take advantage of the smoothness of ball and player trajectories during a typical shot in tennis. Each arc of a shot is functionally encoded by three-degree polynomials in each dimension of 3D space. A shot with 1 bounce consists of 2 arcs and a total of 24 features, a shot with no bounce consists of 1 arc and a total of 12 features. For shots ending with a bounce, the duration of the first arc is fully determined by the z-dimension ending at zero. This leaves the time duration out of the bounce as the only stochastic temporal component to include in the model. Player arcs are encoded as line segments in each dimension in 2D space, which requires 8 additional features. Despite the simplicity of a line to describe the player arcs, our examination of player movement shows that this results in minimal loss of information because the nature of tennis dictates that between any two shots each player either maintains his current position or moves from point A to point B as efficiently as possible. 

\begin{figure}[h]\centering
\begin{tikzpicture}
\tikzstyle{hyper}=[circle, draw =white!80, node distance = 14mm]
\tikzstyle{main}=[circle, minimum size = 12mm, thick, draw =black!80, node distance = 14mm]
\tikzstyle{connect}=[-latex, thick]
\tikzstyle{box}=[rectangle, draw=black!100]
  \node[hyper] (pi) [label=center:$\boldsymbol{\pi}$] { };
   \node[main] (z) [right=of pi,label=center:$z$] {};
   \node[main, fill = black!10] (a) [right=of z,label=center:$\mathcal{A}(\omega)$] {};
  \node[main, fill = black!10] (alpha) [right=of a,label=center:$\mathbf{X}_t$] { };
   \node[main] (mu) [above=of pi,label=center:$\boldsymbol{\mu}$] {};
  \node[main] (sigma) [above=of z,label=center:$\boldsymbol{\Sigma}$] { };
  \node[hyper] (g) [above=of sigma,label=center:$\mathbf{G_0}$] { };
  \path  (pi) edge [connect] (z)
        (z) edge [connect] (a)
        (alpha) edge [connect] (a)
		(mu) edge [connect] (a)
		(sigma) edge [connect] (a)
		(g) edge [connect] (mu)
		(g) edge [connect] (sigma);
  \node[rectangle, inner sep=0mm, fit= (z) (alpha),label=above right:N,xshift=23mm] {};
  \node[rectangle, inner sep=5mm, fit= (z) (alpha),draw=black!100] {};
  \node[rectangle, inner sep=0mm, fit= (mu) (sigma),label=above right:K,yshift=-1mm,xshift=8mm] {};
  \node[rectangle, inner sep=4mm, fit= (mu) (sigma), draw=black!100] {};
\end{tikzpicture}
\caption{Graphical representation of the Gaussian Mixture Model. The $K$ mixture components are drawn from a multinomial distribution. For the $i$th data point of a sample of $N$, $z_i | \lbrace v_1, v_2,\ldots \rbrace \sim Mult(\pi(\mathbf{v}))$ with hyperparameters $v_i | \alpha \sim Beta(1, \alpha)$, $\pi_{i}(\mathbf{v}) = v_i \prod_{j=1}^{i-1} (1-v_j)$, and $G_0$ the the multivariate hyperparameters of the conditional GMM; the stick-breaking representation of an infinite Dirichlet process.}
\end{figure}
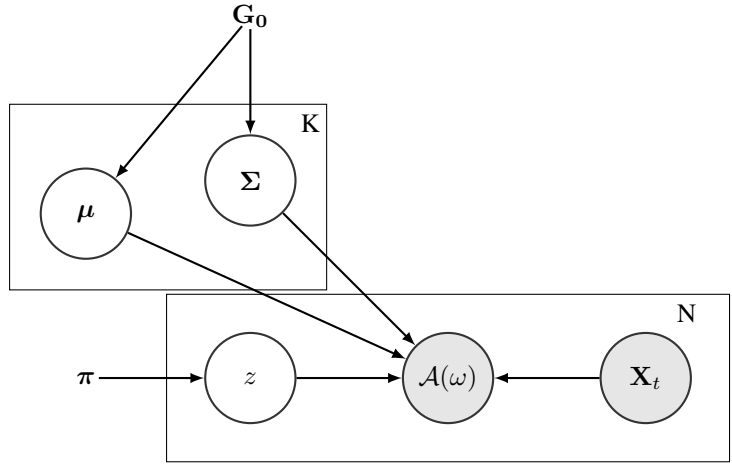

A generative model for the full feature set of $\mathcal{A}(\omega)$ is based on an infinite Bayesian Gaussian mixture model (GMM). The GMM regards each $\mathcal{A}(\omega)$ as belonging to a latent class $z$. Conditional on this class, the functional encoding of the shot event follows a multivariate normal distribution, $\mathcal{A}(\omega) | z \sim MVN(\boldsymbol{\mu}_z, \boldsymbol{\Sigma}_z)$. By using an infinite Dirichlet process prior, the choice of the number and assignment of the mixture components becomes part of the inferential method. Variational inference provides a fast and scalable approach for fitting the GMM by turning the estimation of the target posterior into an optimization problem \citep{blei2006variational}.

\begin{figure}[h] \centering
\includegraphics[width=0.8\textwidth]{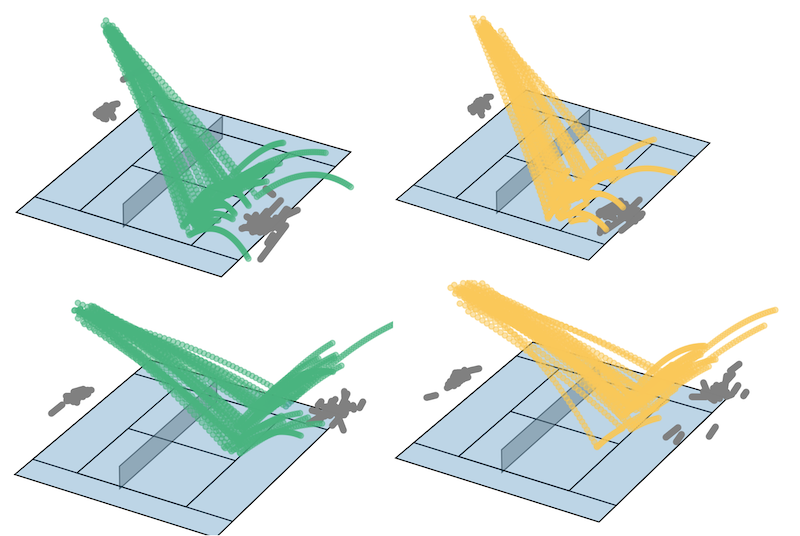}
\caption{3D rendering of ball trajectories and player movement (in grey) for a random sample of men's serves from the 2019 Australian Open (top and bottom left) and an equal sample of shots from the GMM for serves (top and bottom right).}
\label{fig:real-fake}
\end{figure}

We use the GMM implementation provided by the scikit-learn library in python \citep{scikit-learn}, setting the upper bound for the mixtures of 20 and using a full covariance structure. Seperate generative models were fit for shots with and without a bounce within each of three major shot types: serves, serve returns, and rally shots. All training was based on three years of tracking data from the Australian Open, with 20\% of the data reserved for out-of-sample testing. The out-of-sample log-likelihood showed good agreement between real and `fake' shot events across all of the shot types. An illustration of the generative performance is given in Figure~\ref{fig:real-fake} where a 3D rendering of generated and observed men's serves to the Ad and Deuce court demonstrates that the output of the GMM is virtually indistinguishable from actual shots.

\subsection{Conditional Generation}

We now consider how to update the generative distribution of shot events, $\mathcal{A}(\omega)$, given ball and player positions up to some time $t_0$. Without a loss of generality, we will consider the case in just one dimension of the first arc of a shot, $\mathcal{A}(\omega) = (\theta_0, \theta_1, \theta_2, \theta_3)$, where the parameters describe the movement of the ball in the x-dimension,

\begin{equation}
f_{x}(t) = \theta_0 + \theta_1 * t + \theta_2 * t^2 + \theta_3 * t^3
\label{eq:poly}
\end{equation}

\noindent Suppose that at time $t_0$ we observe the location $f_{x}(t_0)$. The key insight for conditioning $\mathcal{A}(\omega)$ on this information is to recognize that, if time is treated as fixed, $f_{x}(t_0)$ is a linear combination of the parameters $\mathcal{A}(\omega)$. Letting $\mathbf{C} = \left(\begin{array}{cccc} 1 &  t &  t^2 &  t^3 \end{array}\right)$ be the $1 \times 4$ set of constant time components for the polynomial and $\mathcal{A}(\omega)$ the $4 \times 1$ vector of $\theta$ coefficients in Eq.~(\ref{eq:poly}), every observation $f_{x}(t)$ can be expressed as $\mathbf{C} \mathcal{A}(\omega) = f_{x}(t)$. 

Properties of MVN tell us that, if $\mathcal{A}(\omega) \sim MVN \left(\boldsymbol{\mu},  \boldsymbol{\Sigma} \right)$, any linear combination of the form $\mathbf{C} \mathcal{A}(\omega) + \mathbf{b}$, for constants $\mathbf{C}$ and $\mathbf{b}$, will be distributed as  $MVN \left( \mathbf{C} \boldsymbol{\mu} + \mathbf{b}, \mathbf{C} \boldsymbol{\Sigma} \mathbf{C}' \right)$. From this result, we can easily construct the joint likelihood of $\mathcal{A}(\omega)$ and $\mathbf{C} \mathcal{A}(\omega)$,

\begin{equation}
\left(\begin{array}{c} \mathcal{A}(\omega) \\ \mathbf{C} \mathcal{A}(\omega)\end{array}\right)  \sim MVN\left(\left(\begin{array}{c}\boldsymbol{\mu} \\ \mathbf{C} \boldsymbol{\mu} \end{array}\right), \left( \begin{array}{cc} \boldsymbol{\Sigma} &  \boldsymbol{\Sigma} \mathbf{C}'\\    \mathbf{C} \boldsymbol{\Sigma}  & \mathbf{C} \boldsymbol{\Sigma} \mathbf{C}'  \end{array} \right) \right)
\end{equation}

\noindent Well-known properties of MVN random variables also provide the solution to the conditional density, $\mathcal{A}(\omega) | \mathbf{C} \mathcal{A}(\omega) = f_{x}(t)$, which is an MVN with mean $\boldsymbol{\mu} + \boldsymbol{\Sigma} \mathbf{C}' (\mathbf{C} \boldsymbol{\Sigma} \mathbf{C}') ^{-1} (f_x(t) - \mathbf{C} \boldsymbol{\mu})$ and variance,

\begin{equation}
\boldsymbol{\Sigma} - \boldsymbol{\Sigma} \mathbf{C}' (\mathbf{C} \boldsymbol{\Sigma} \mathbf{C}') ^{-1} \mathbf{C} \boldsymbol{\Sigma}.
\label{eq:mvn_cond}
\end{equation}

The $\mathbf{C}$ matrices and the conditional formula in Eq.~(\ref{eq:mvn_cond}) can be constructed for the ball and player positioning for any number of data points in time. The final step to completing the full conditioning of the GMM is the updating of the mixture weights. Let $\mathbb{P}(\mathbf{X}_t|z)$ be the MVN update given observed data $\mathbf{X}_t$ for the mixture component $z$. If each $z$ has the prior weight $\mathbb{P}(z)$ the update once $\mathbf{X}_t = \mathbf{x}_t$ is observed is,

\begin{equation}
\mathbb{P}(z | \mathbf{X}_t = \mathbf{x}_t) = \frac{\mathbb{P}(\mathbf{X}_t = \mathbf{x}_t|z)\mathbb{P}(z)}{\sum_z \mathbb{P}(\mathbf{X}_t = \mathbf{x}_t | z) \mathbb{P}(z)}
\end{equation}

\begin{figure} \centering
\includegraphics[width=0.5\textwidth]{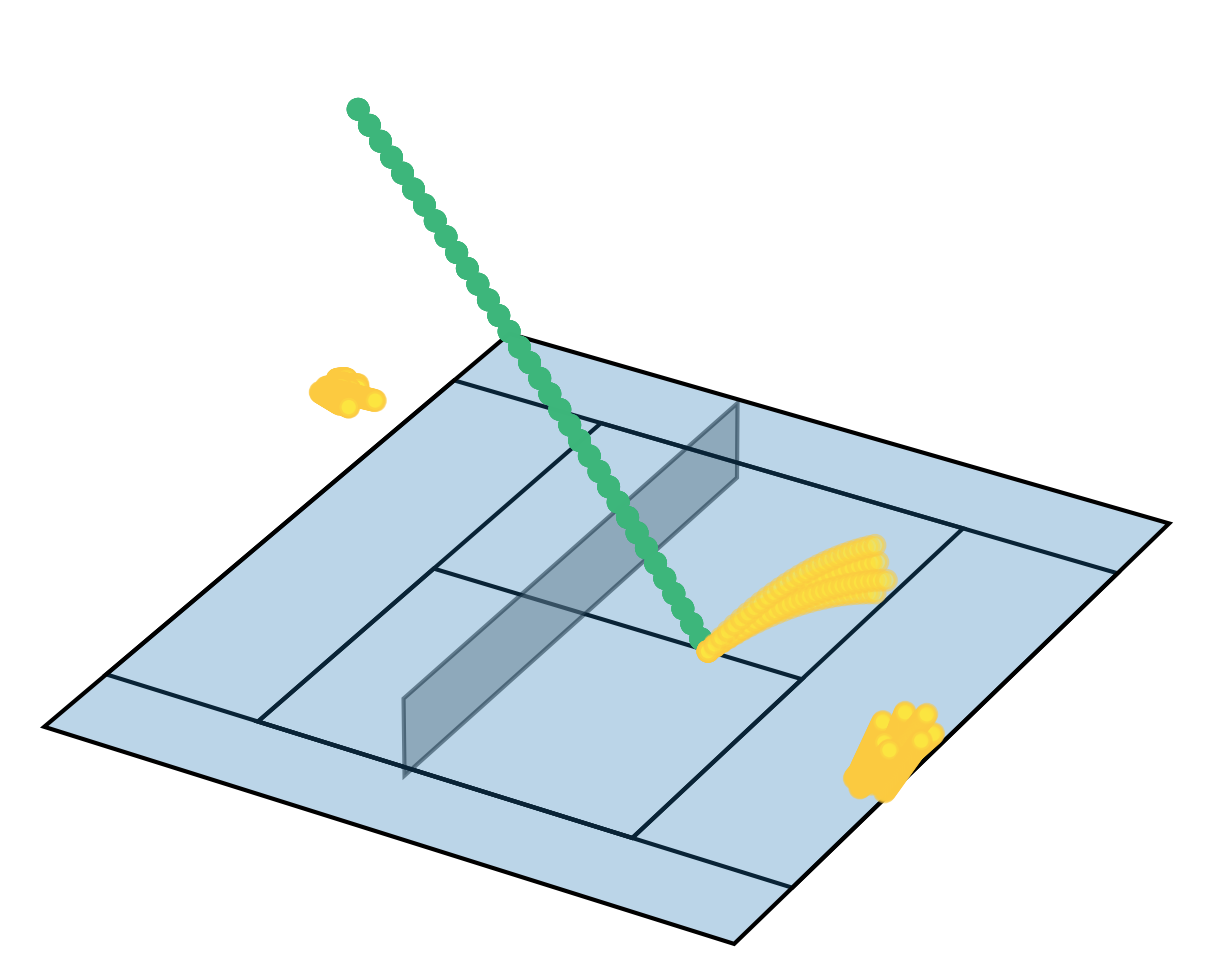}
\caption{Example of generative conditioning for a down the T serve to the Ad court with conditioning on the arc of the trajectory into the bounce (in green).}
\label{fig:conditioned}
\end{figure}

\noindent with each component of the sum being a simple likelihood calculation for the MVN. Figure~\ref{fig:conditioned} provides an illustration of 10 randomly sampled shots from the conditioned GMM for a serve to the Ad court conditioned on all of the shot trajectory into the bounce.

\subsection{Outcome Prediction}

When we have the ability to generate possible futures of a shot event from any time into the shot, we can use all of the information contained in each projected shot event to derive predictive features for outcomes of interest. For our measure of shot value, we focus on the outcome of a shot that earns a point for the player who made it, which we define as shots that are clean winners or induce an error from the opponent. Predictive models to distinguish winning shots from shots in play were developed using  generalized additive models (GAMs); shots ending in errors can be identified by the bounce location of the shot and were automatically assigned a value of zero. Guided by prior shot prediction work, we derived multiple features based on the speed, height, and bounce location of the ball \citep{wei2013predicting}. We also considered multiple features of the opponent, including their handedness, location on court, and the implied distance and speed they would need to travel to return the ball in a good position, defining a `good position' as the location of the ball when it first reaches 1 meter in height out of the bounce. One of the strengths of the GAM is its ability to model spatial effects non-parametrically through the use of smoothing functions \citep{wood-jasa}. We applied thin plate splines for all the spatial features and considered univariate as well as bivariate smooths. Selection among different choices of smoothing functions was handled as part of the model fit through the use of a penalized likelihood. All models with fit with the \texttt{mgcv} package in R \citep{wood-gam}.

\begin{table*} \centering
    \caption{Hold-out sample performance of GAM models for predicting a winning shot}
    \begin{tabular}{ll ccc} \hline
        Shot Type & Shot Subset & Sample Size & Precision & Recall  \\ \hline
        Serves &  Men's In Play  & 3,817    & 0.82 &  0.98 \\ 
            & Men's Won Point & 1,253 & 0.85 & 0.35 \\
            & Women's In Play  & 2,846  & 0.86 & 0.88 \\
            & Women's Won Point  & 647 & 0.88 & 0.30 \\ \noalign{\smallskip}
                   Rally &  Men's In Play   & 8,312   & 0.93 &  0.98 \\ 
            & Men's Won Point  & 1,701 & 0.86 & 0.66 \\
            & Women's In Play  & 4,252  & 0.92 & 0.98 \\
            & Women's Won Point  & 1,100 & 0.92 & 0.68 \\ \noalign{\smallskip}
             Rally (+Incoming) &   Men's In Play   & 8,312   & 0.94 &  0.98 \\ 
            & Men's Won Point  & 1,701 & 0.87 & 0.68 \\
            & Women's In Play  & 4,252  & 0.92 & 0.98 \\
            & Women's Won Point  & 1,100 & 0.91 & 0.68 \\  \hline
    \end{tabular}
    \label{tab:gam-fit}
\end{table*}

Separate models were fit for serves and rally\footnote{Counting serve returns and all subsequent shots as rally shots} shots. Because rally shots are a response to an incoming shot, we also fit models that included the dominance features of the incoming shot relative to the current shot as defined in \citep{wei2016thin}. Log-loss, calibration, precision and recall were the main metrics used to assess the performance of the models in the 20\% hold-out sample. All of the models had good calibration, with each decile bin having a mean difference of $\leq$0.4 percentage points the decile's observed percentage of winning shots. Log-loss was higher for serve predictions compared to rally shots (0.44 compared to 0.19), owing to lower recall for predicted winning shots (Table~\ref{tab:gam-fit}) which we believe is driven by greater ambiguity in the classification of `forced' and `unforced' errors on serve returns by current providers of point-by-point scoring feeds and our choice to only consider shots inducing `forced` errors as winning shots. The addition of dominance features for the incoming shot did not result in a meaningful improvement in the predictive performance for outcomes of rally shots.

\section{ESV}

\begin{figure}[h] \centering
\includegraphics[width=0.8\textwidth]{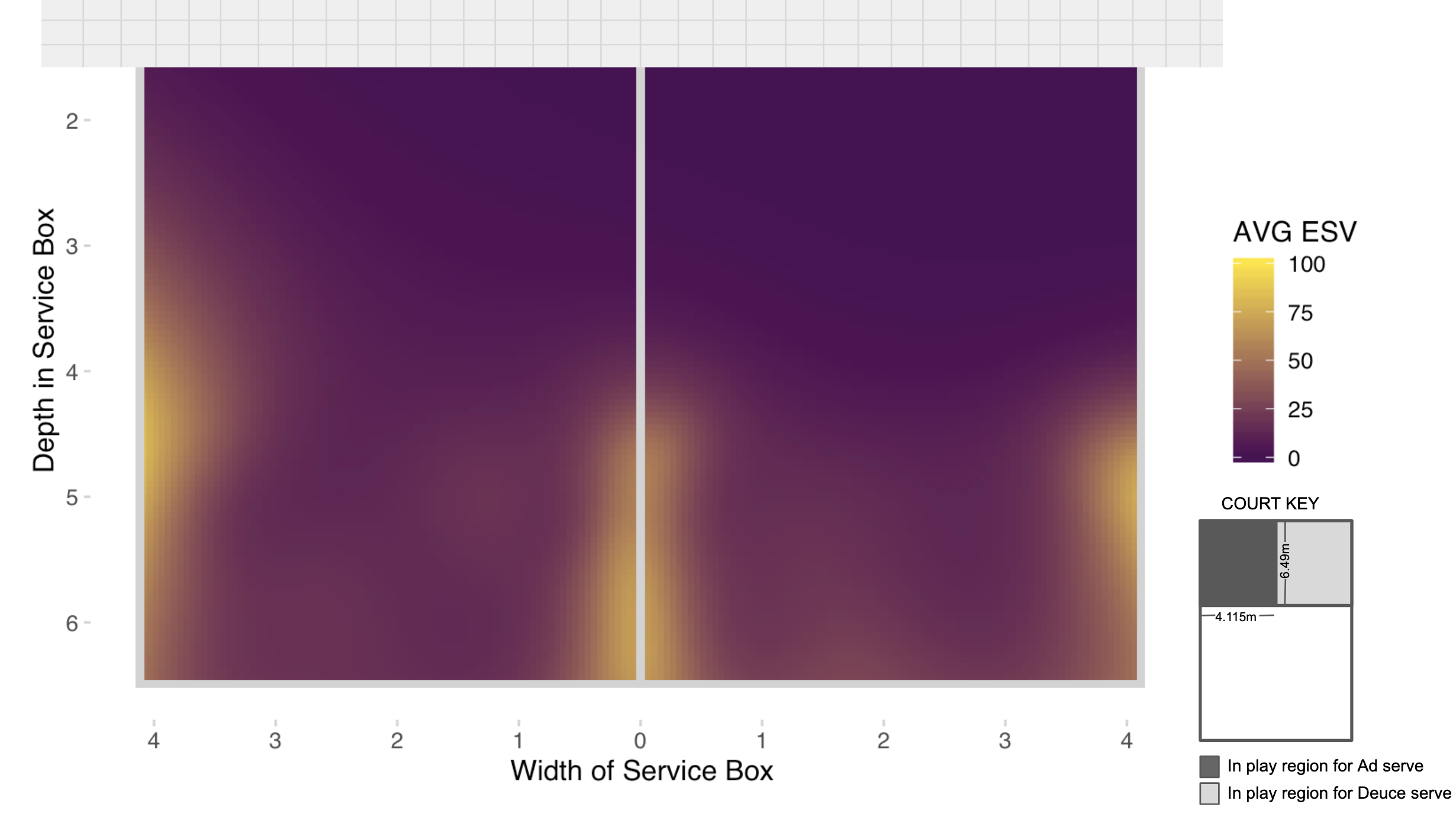}
\caption{Average expected shot value (ESV) of men's first serves at the 2019 US Open by their landing location.}
\label{fig:esv-serve}
\end{figure}

Bringing all of the above ingredients together gives us the ability to study aspects of the game that have long been out-of-reach of statistical analysis in tennis. We can use ESV, for example, to identify the spatial characteristics of the most highly valued serves. A heat map of the ESV of men's first serves from the 2019 US Open shows how critical angles are for an effective serve, the most valued regions being far wide or down the center (Figure~\ref{fig:esv-serve}). While these characteristics have generally been known for some time, ESV gives us the ability to examine whether there is difference in value between a wide and down-the-T serve. By looking at the proportion of shots over $>50$\% ESV that are within 1m of the service box wide compared to 1m of the center line, we find a slight advantage for wide serves (11\% vs 8\%). This can be partly driven by short serves out wide to the Ad court being especially difficult for right-handed opponents, whereas the T serve does not appear to be any more difficult when directed to the backhand versus the forehand side.

\begin{figure}[h]
     \centering
     \begin{subfigure}[t]{\textwidth}
         \centering
    \includegraphics[width=0.8\textwidth]{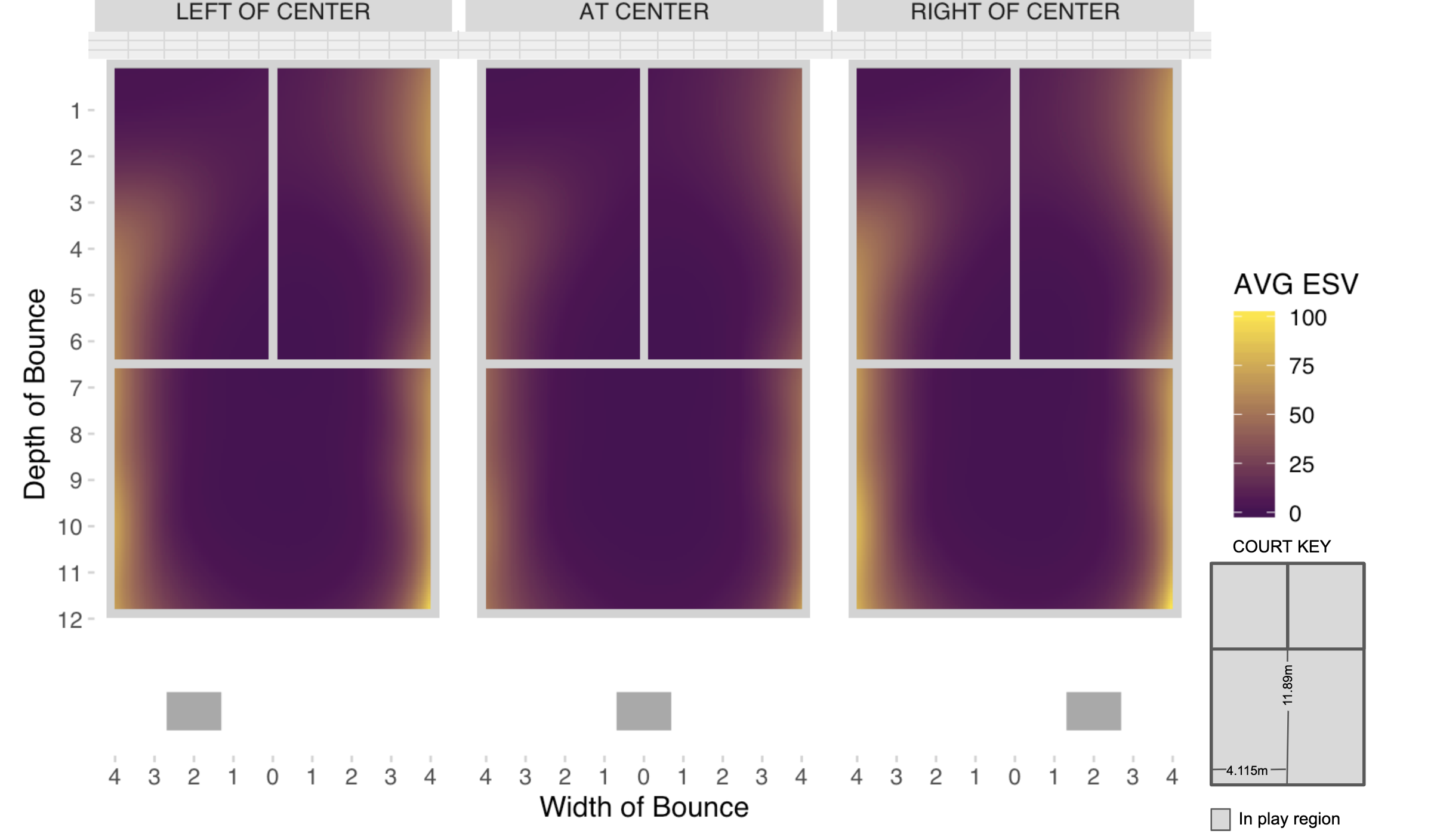}
    \caption{Men's forehand shots}
    \label{fig:esv-fh}
    \end{subfigure}
    \begin{subfigure}[t]{\textwidth}
         \centering
    \includegraphics[width=0.8\textwidth]{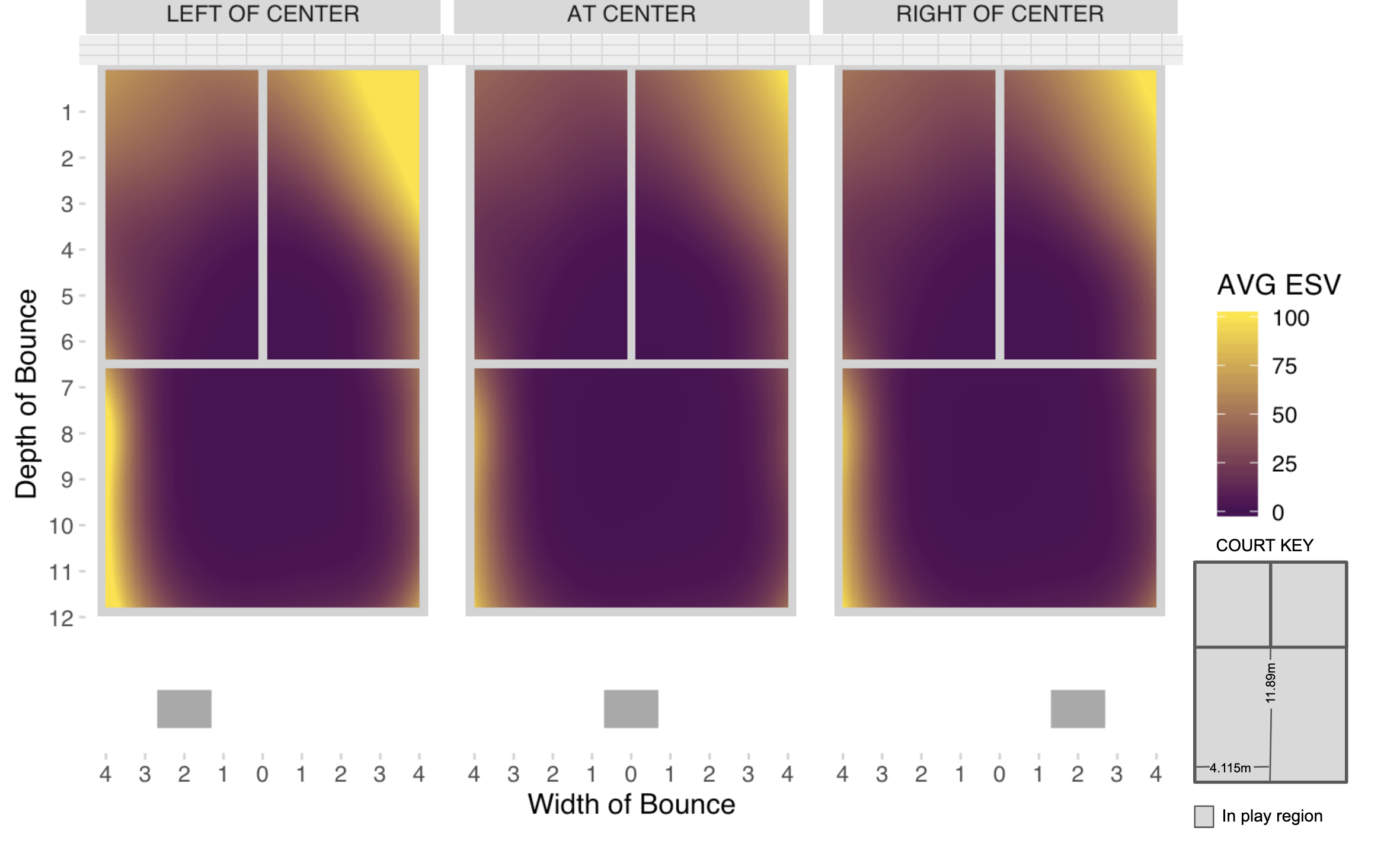}
    \caption{Men's backhand shots}
    \label{fig:esv-bh}
    \end{subfigure}   
    \caption{Average expected shot value (ESV) of rally shots at the 2019 US Open. Points corresponding to the landing location of the forehand and the grey box indicates the location of the opponent when the shot was made. Only shots within two shots of the point-ending shot are included.}
    \label{fig:esv-strokes}
\end{figure}

Using a similar visualization of the spatial location of shots that are within 2 shots from the last shot of the rally can highlight the most valuable landing locations of winning shots as well as shots that setup the win. We also consider how the spatial value varies with the location of the opponent when the shot comes off the racquet, considering an opponent positioned behind the baseline but differing in their location out wide. In the case of forehand shots (Figure~\ref{fig:esv-fh}), we see the region nearest to the lines out wide being the highest valued in all situations. However, there are notable differences in the distribution of the ESV in these regions depending on the opponent's position. A centrally located opponent is the most likely to neutralize the forehand shot as this situation has the fewest regions of high ESV. When opponents are further out wide, shots within 1m of a sideline have a strong probability of winning the point; the strength increasing somewhat in the case of a right-handed opponent out to the right with more area for the backhand to cover. We also find less dependency on the width of the shot if a ball is hit extremely short (within 4 meters of the net) though the region that is within 1m of the centre still remains a region of low ESV.

Spatial heat maps of backhand shots show a clear difference in the value of the left and right corners of the baseline, the left corner having the greater value across the different opponent positions considered (Figure~\ref{fig:esv-bh}). This highlights the effectiveness of down-the-line backhand shots directed to the backhand side of the opponent. The maps also show the top right corner within a few meters of the net as a region of high ESV. This region is the result of backhand drop shots that US Open players tended to direct to the forehand side of the court. Although this is a high-valued shot whenever the opponent is beyond the baseline, it is most effective when players are left of center and would have the most distance to travel.

\section{ESV Metrics}

With the decoupling that ESV allows, we can derive a number of novel performance statistics. In what follows, we present three ESV metrics: VAST, ShotIQ, and VACC. Each are applied to tracking data from the 2019 US Open. 

\subsection{VAST}

VAST, for value added with shot taking, is the shot value an impact player achieves against an average player in the same situation. For a given shot, the encoding $\mathcal{A}(\omega)$ can be partitioned into the variables that are specific to the shot and player who made the shot, $\mathcal{A}_{\mathcal{S}}(\omega)$, and the variables that are specific to the receiver of the shot, $\mathcal{A}_{\mathcal{R}}(\omega)$. The calculation of VAST involves the marginalization of the ESV over the receiver variables,

\begin{equation}
\mbox{VAST} = \int_{\mathbb{R}} \mathbb{P}(W(\omega) | \mathcal{A}_{\mathcal{S}}(\omega) = S, \mathcal{A}_{\mathcal{R}}(\omega) = R) \mathbb{P}(\mathcal{A}_{\mathcal{R}}(\omega) = R) dR,
\label{eq:vast}
\end{equation}

\noindent which can be readily estimated with Monte Carlo techniques. The result is an estimate of the ESV for the observed actions of the impact player that would be expected on average, regardless of the behavior of the receiving player. In this way, VAST isolates the value of a player's shot selection and execution. 

\begin{figure}[h]
     \centering
     \begin{subfigure}[t]{0.45\textwidth}
         \centering
         \includegraphics[width=0.8\textwidth]{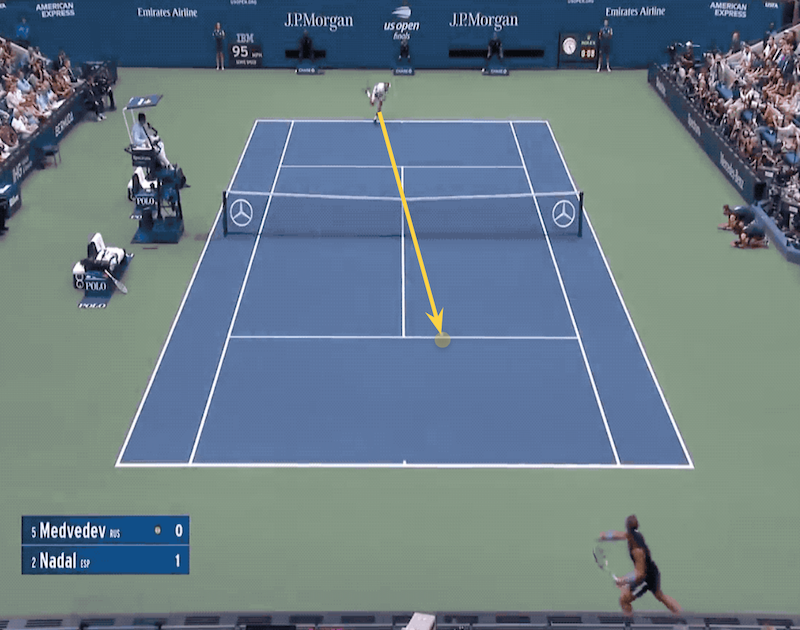}
         \caption{Daniil Medvedev serving against Rafael Nadal at the 2019 US Open men's final; a down-the-T serve that is returned in play by Nadal.}
         \label{fig:vast1}
     \end{subfigure}
     \hspace{0.5em}
     \begin{subfigure}[t]{0.5\textwidth}
         \centering
         \includegraphics[width=0.8\textwidth]{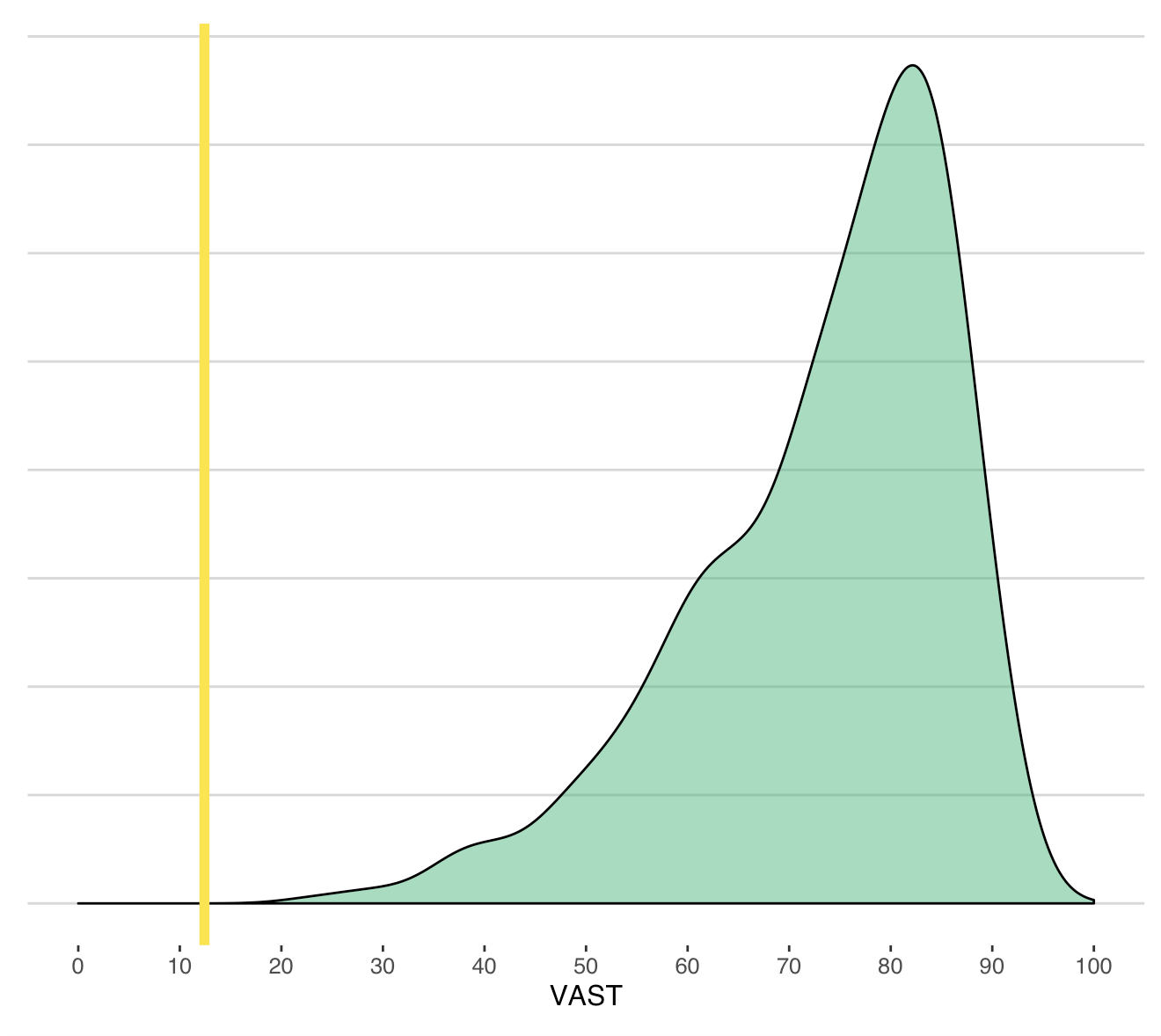}
         \caption{Marginalized ESV distribution of Medvedev's serve, integrating over receiver behavior. Vertical line is the predicted chance the serve wins the point given Nadal's actual positioning and movement.}
         \label{fig:vast2}
     \end{subfigure}
        \caption{Example of how receiver behavior can impact the returnability of a serve as adjusted for by VAST.}
        \label{fig:vast_example}
\end{figure}

While outcomes on the serve\textemdash like the frequency of aces\textemdash are often attributed solely to the ability of the server\footnote{See for example the `Serve Rating' of the ATP, which rates players on their serve outcomes without any adjustment for receiver behavior \url{https://www.atptour.com/en/stats/leaderboard?boardType=serve}}, the VAST metric shows that receivers can have a considerable impact on whether a serve wins the point outright. Consider the serve in Figure~\ref{fig:vast_example}, the first serve hit by Daniil Medvedev to Rafael Nadal in the 2019 men's US Open final. That down-the-T serve moving at 121 mph has a marginal distribution with a VAST of 75\% and puts 95\% of possible outcomes with a predicted win chance greater than 50\%. Against any other opponent this would be a near certain point won for Medvedev but against Nadal, considering the depth of his position, his read on the serve, and the time he has to close the gap between himself and the ball as it comes out of the bounce, the chance that the serve is a winning shot is just 1 in 10. The fact that an opponent can have as substantial an influence as this on the outcome of any shot, let alone the serve, affirms the need for statistics like VAST that can separate player and opponent actions when evaluating outcomes in a tennis point. 

\begin{figure}[h]
     \centering
     \begin{subfigure}[t]{0.7\textwidth}
         \centering
         \includegraphics[width=0.8\textwidth]{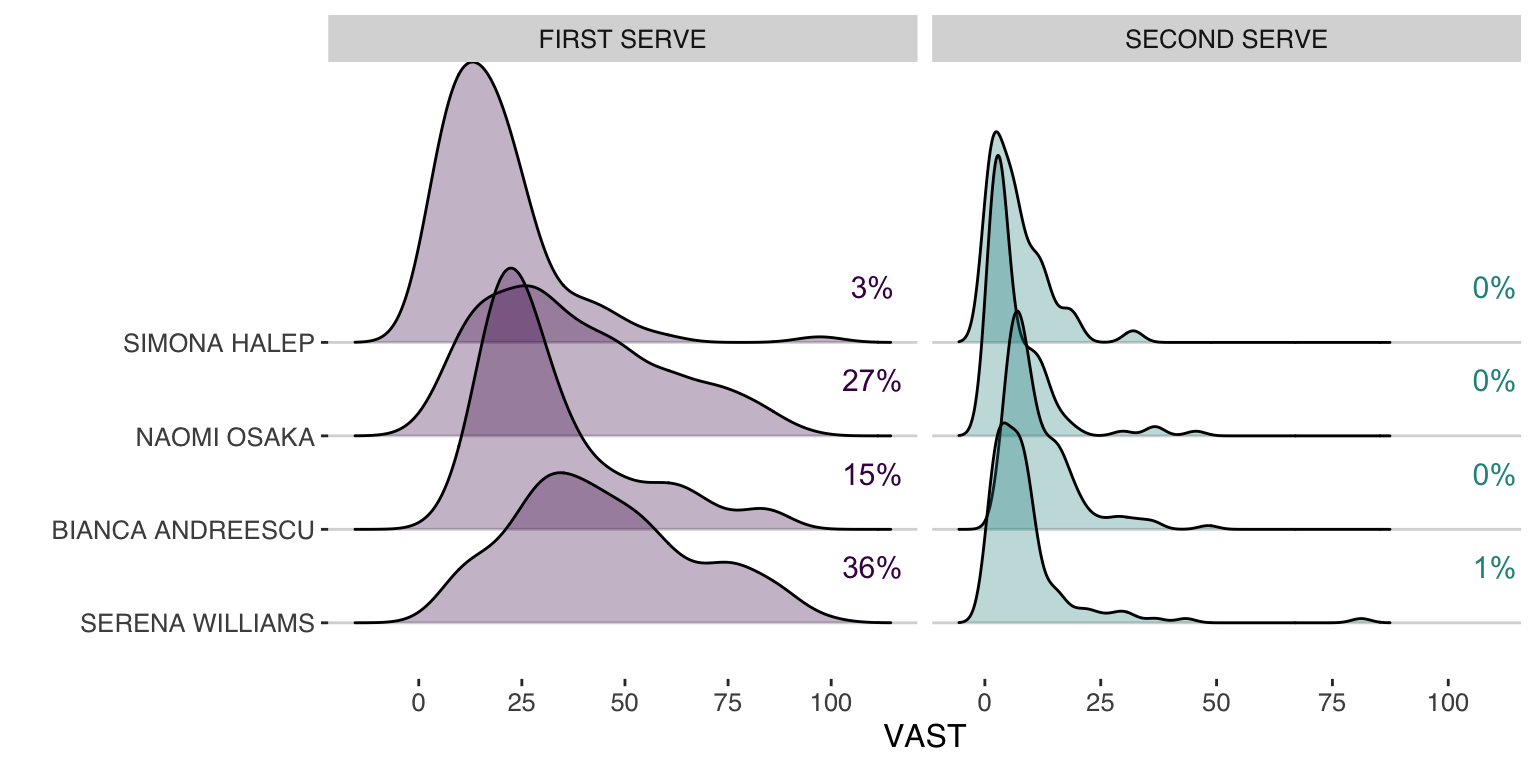}
         \caption{VAST for first and second serves}
         \label{fig:vast-serve}
     \end{subfigure}
     \vspace{0.5em}
     \begin{subfigure}[t]{0.8\textwidth}
         \centering
         \includegraphics[width=0.8\textwidth]{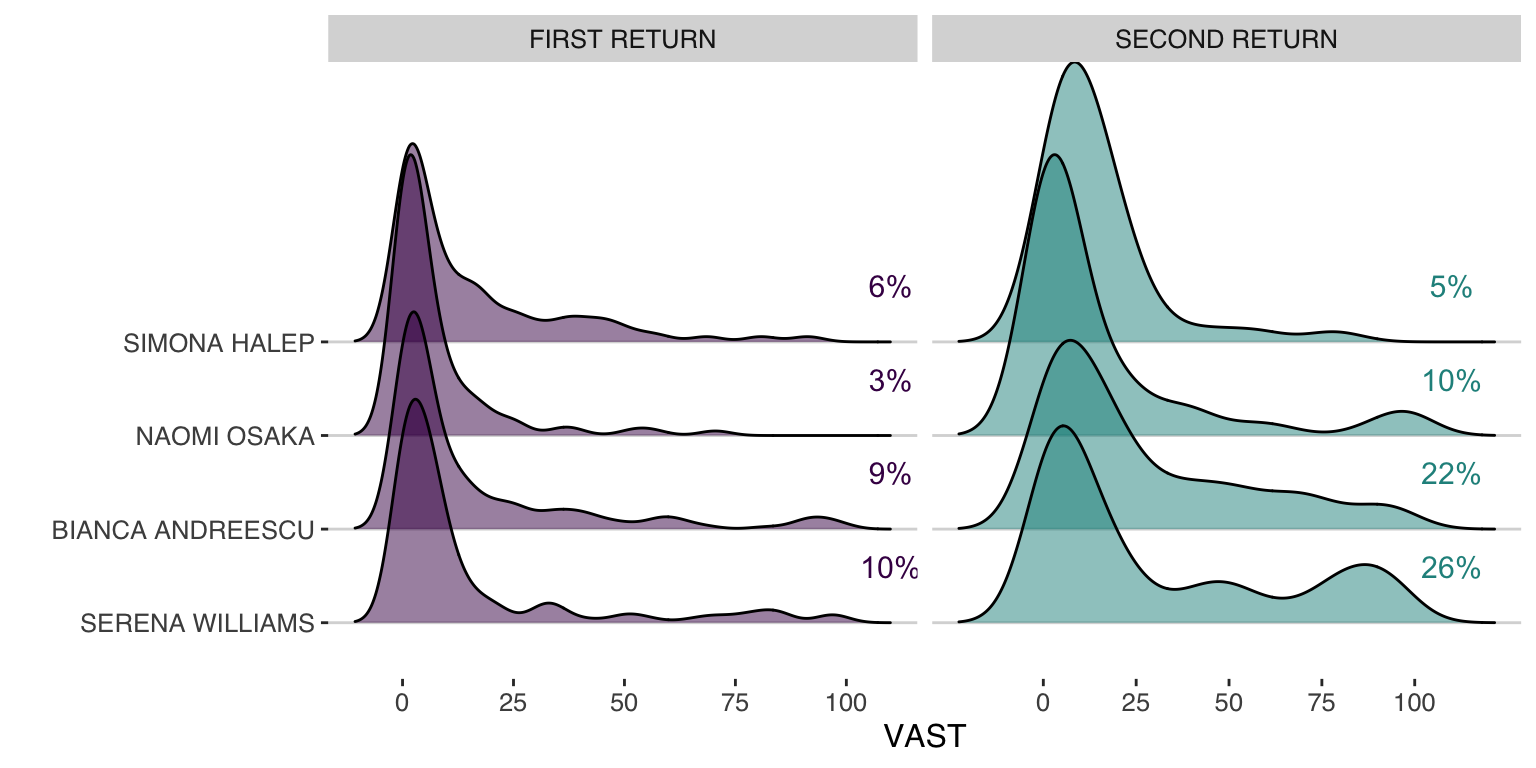}
         \caption{VAST for first serve returns and second serve returns}
         \label{fig:vast-servereturn}
     \end{subfigure}
     \vspace{0.5em}
    \begin{subfigure}[t]{0.7\textwidth}
         \centering
         \includegraphics[width=0.8\textwidth]{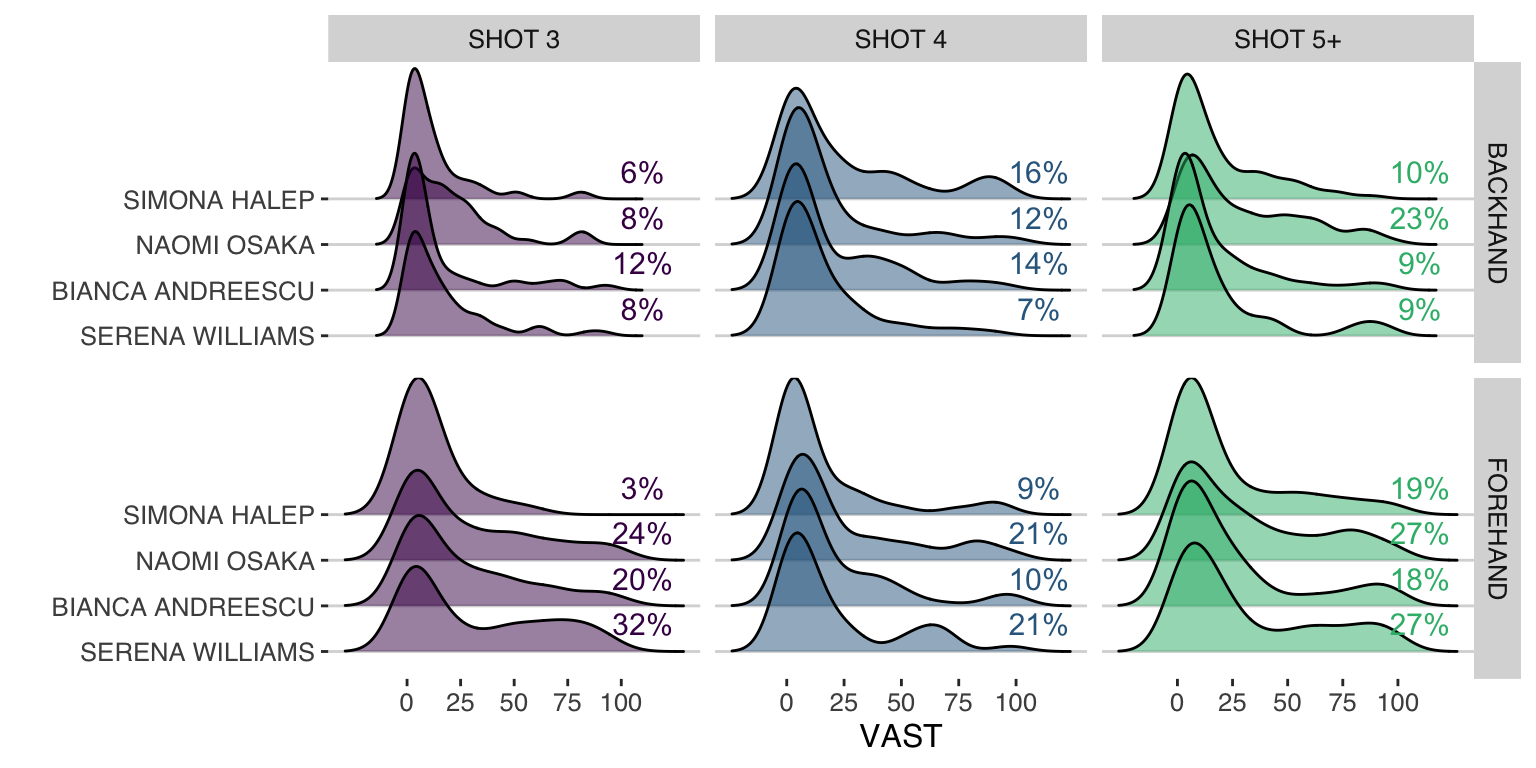}
         \caption{VAST for forehand and backhand shots grouped by shot number, `3' corresponding to a server's first rally shot, `4' the receiver's first rally shot, and `5+' all other rally shots.}
         \label{fig:vast-rally}
     \end{subfigure} 
        \caption{VAST at the 2019 US Open by shot type for Top 4 WTA players. The annotated numbers show the area of the density for a VAST of 50\% or greater.}
        \label{fig:vast_wta}
\end{figure}

Detailed summaries of VAST by shot type can provide unique insights into the technical strengths of a player's game. Figure~\ref{fig:vast_wta} shows such a breakdown for all shots played at the 2019 US Open for the 4 highest ranked women's players at the time of this writing. The dominance of Serena Williams' first serve is made clear by the density of her shots with VAST values of 50\% or more, suggesting that Williams won more than 1 or every 3 points on her first serve alone at the US Open. While the second serve is a safe shot for all of the players, it is interesting to see that Andreescu and Williams have very similar VAST distributions for this shot, both more aggressive relative to Halep and Osaka. The similarity holds for serve returns as well, where both Williams and Andreescu have the most attacking serve returns, each reaching VAST scores over 50\% for more than 1 in every 5 serve returns on second serves. In rallies, Williams is the master of the one-two pattern on serve, having the strongest VAST distribution on the third shot when hitting with her forehand. Osaka has a Serena-like game on her forehand while the strength of her backhand comes out most in longer rallies. Andreescu's patterns closely parallel Williams', lagging behind the difficulty of Williams' forehand 3rd shot but excelling on the difficulty of her backhand in short and long rallies. 

\subsection{Shot IQ}

Shot IQ attempts to decouple smart shot choices from the quality of a shot's execution, highlighting when a player went for the tactically `right' shot even if the execution was wanting. The estimation of Shot IQ proceeds in a similar way as with VAST but the partitioning of the shot event encoding,  $\mathcal{A}(\omega)$, is different. For Shot IQ, the variables included among the fixed features, $\mathcal{A}_{\mathcal{S}}(\omega)$, are the the location of the player and the location of the opponent when the shot leaves the racquet, as well as the eventual bounce location of the shot. These are the minimal features needed to describe the positional configuration of the shot and the players. All remaining variables, such as the speed and shape of the shot and the speed and movement of the players, make up $\mathcal{A}_{\mathcal{S}}(\omega)$ and are the variables that Shot IQ integrates out as in Eq.~(\ref{eq:vast}).

\begin{figure}[h]
     \centering
     \begin{subfigure}[t]{0.43\textwidth}
         \centering
         \includegraphics[width=0.8\textwidth]{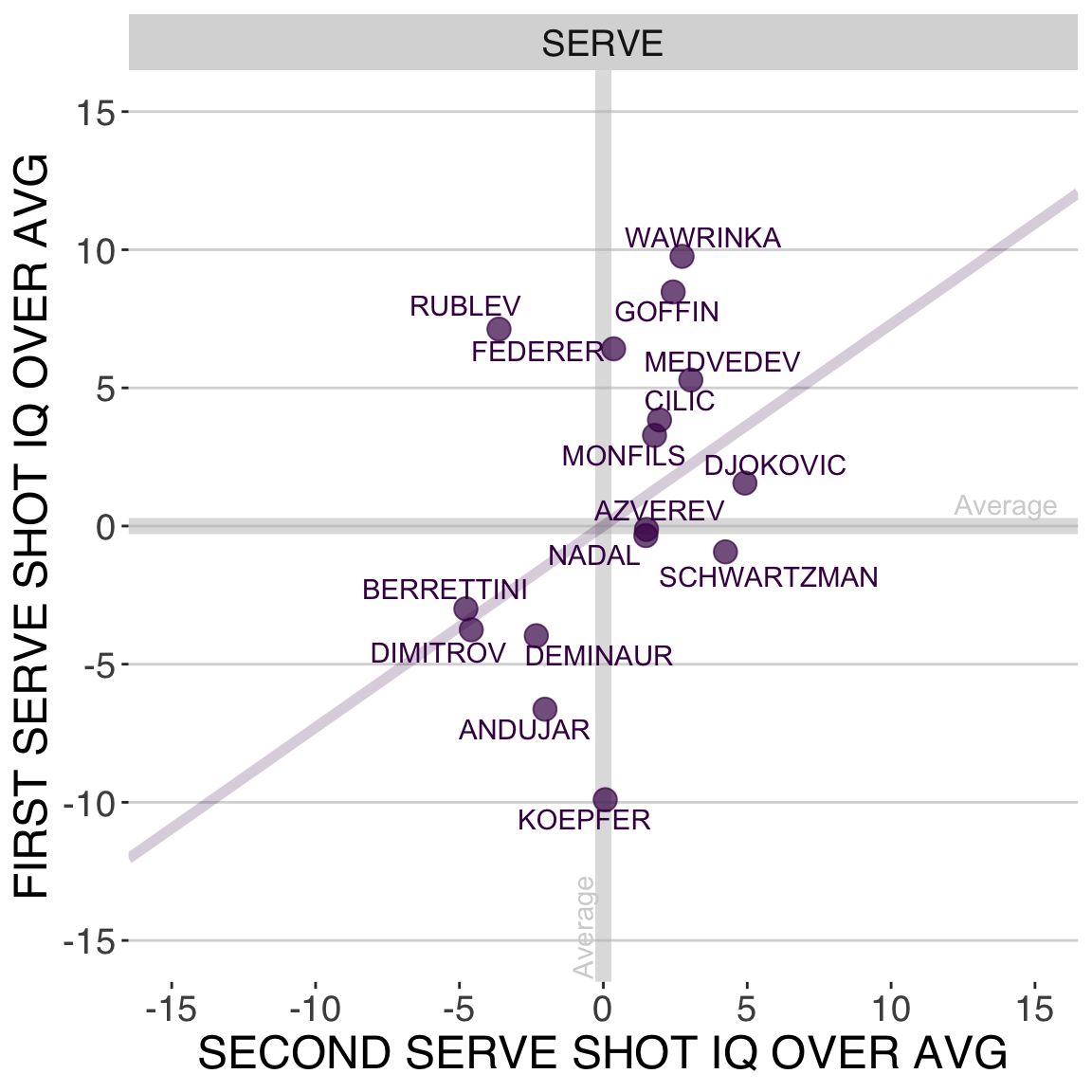}
         \caption{Mean Shot IQ over average on first and second serves}
         \label{fig:shotiq1}
     \end{subfigure}
     \hspace{0.1em}
      \begin{subfigure}[t]{0.43\textwidth}
         \centering
         \includegraphics[width=0.8\textwidth]{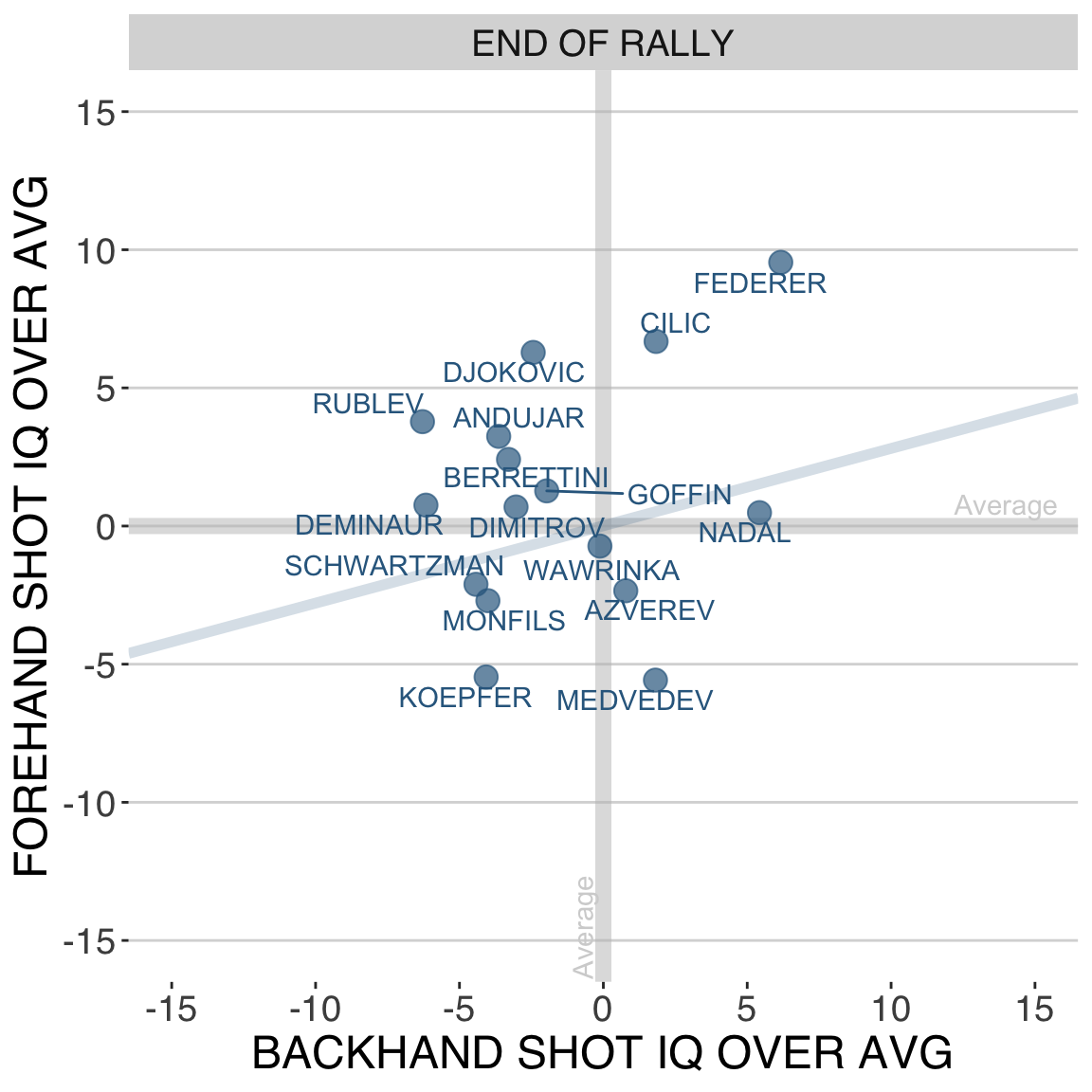}
         \caption{Mean Shot IQ over average for rally shots within 2 shots of the point-ending shot}
         \label{fig:shotiq3}
     \end{subfigure} 
        \caption{Shot IQ summaries at the 2019 US Open for players who reached the round of 16}
        \label{fig:shotiq}
\end{figure}

When we apply the Shot IQ to the male players who reached the Round of 16 at the 2019 US Open, we find that the performance of this group ranged between -10 and +10 percentage points on first serve and -5 and +5 on second serve compared to the average at the event (Figure~\ref{fig:shotiq}). In general we see a positive association between first and second serve Shot IQ. Five players, including Stan Wawrinka and countrymen Roger Federer, have an especially high Shot IQ on first serve, while Novak Djokovic and Diego Schwartzman earn the highest scores on second serve. Schwartzman's superior performance on second serve Shot IQ demonstrates the ability of this stat to separate decision-making about the location of shot from execution quality as, at just 5ft 7in, Schwartzman is usually ranked below average on most service stats.

In Figure~\ref{fig:shotiq3}, we focus on the final shots of a rally where, as point-ending shots, we expect them to be a more difficult subset of rally shots. In contrast to first serves, there is less somewhat less variance between players on rally-ending forehand and backhand Shot IQ, the range being -5 to +10 percentage points over average. We also see a much weaker relationship between forehand and backhand Shot IQ compared to first and second serves, which could be explained by the greater difference in strategy and technique between forehand and backhand shots. In fact, only 3 of the Round of 16 players were found to have above average Shot IQ on forehand and backhand: Roger Federer, Marin Cilic, and Rafael Nadal (all former winners of the US Open; Nadal the champion in 2019). Even among these three, Federer excels in having the highest forehand and backhand Shot IQ, confirming his status as one of the smartest players on tour when it comes to shot selection. 

Like Federer and Cilic, Novak Djokovic had a Shot IQ on the forehand that was more than 5 points greater than the average male player at the US Open. Yet Djokovic trailed slightly behind the average when it came to the placement of his backhand. This runs counter to expert opinion that ranks Djokovic's backhand as the best two-handed backhand in the current game \footnote{\url{https://www.nytimes.com/2018/11/09/sports/tennis/who-has-the-best-shots-in-mens-tennis.html?}}, which raises the question of whether this assessment is owing to the superior execution of the shot or if the US Open wasn't representative of Djokovic's typical backhand performance.

\subsection{VACC}

VACC, for value added due to court coverage, flips the perspective from the player making the shot to the receiving player. The purpose of VACC is to attribute value to a player for their defense, which in tennis is driven by a player's ability to maintain a good position and move well when retrieving shots. Suppose that the actual movement of the opponent during a given shot event is $\mathcal{A}_{\mathcal{R}}(\omega) = R$, the shot-level estimate of VACC is

\begin{equation}
\mbox{VACC} = \mbox{VAST} - \mathbb{P}(W(\omega) | \mathcal{A}_{\mathcal{S}}(\omega) = S, \mathcal{A}_{\mathcal{R}}(\omega) = R).
\label{eq:vacc}
\end{equation}

\noindent In words, Eq.~(\ref{eq:vacc}) measures the gap between the difficulty of a shot against an average receiver and the difficulty of the same shot against the current receiver. The more a receiver is able to neutralize a shot through their court coverage, the more positive their VACC; while a receiver who is less effective in returning shots than an average player would have a negative VACC.

\begin{figure}[h]
    \centering
    \includegraphics[width=0.8\textwidth]{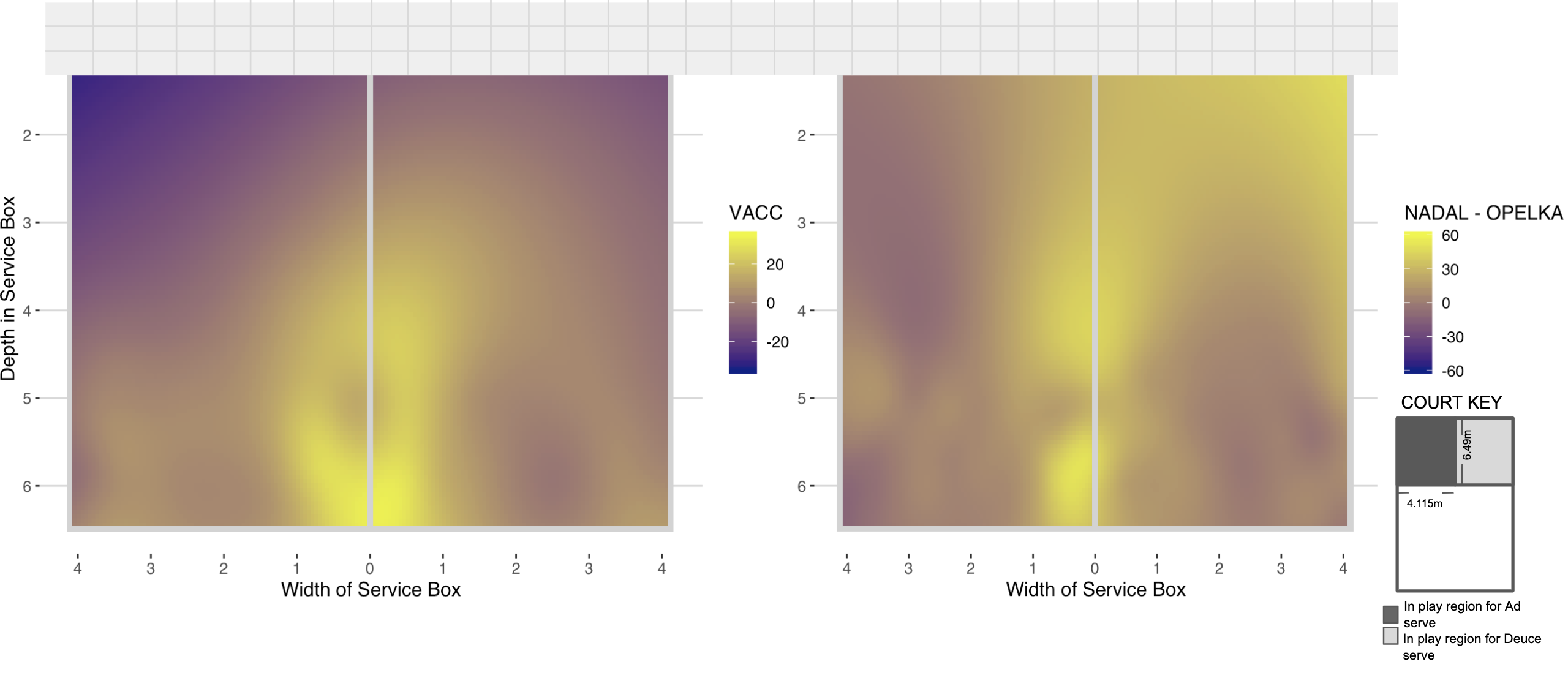}
    \caption{Heat map of VACC for Rafael Nadal when receiving first serve (left panel) and the difference between the VACC of Nadal and Reilly Opelka (right panel) based on bounce location of serve for all first serves received at the 2019 US Open}
    \label{fig:vacc}
\end{figure}

Analysis of VACC statistics at the US Open shows that Rafael Nadal is highly effective in neutralizing the strength of his opponent's serve. Across all of his matches, Nadal achieved a VACC of 7.5\% when facing first serves, the most powerful shots in tennis. This means that Nadal reduced the chance of the server winning a point on their serve by 7.5 percentage points on average, owing to his positioning and his movement. For 1 in 7 first serves Nadal's defensive skill was as high as 30\% or more. 

It is generally accepted that extreme height, while being an advantage for serve, can be a liability when it comes to defensive skill in tennis, as the tallest professional players tend to move less effectively than competitors of average height. We can see evidence of this phenomenon by contrasting the VACC maps between 6ft 11inch Reilly Opelka and 6ft Rafael Nadal when receiving first serve (Figure~\ref{fig:vacc}, right panel). Overall, Nadal has a +15 average VACC over Opelka. In addition to the down-the-T regions where we have already seen Nadal's coverage excels compared o the average player, we can observe some asymmetries in the highest VACC differences between the Ad and the Deuce court. The areas where the gap between Nadal and Opelka is greatest on the Ad court is along the centre line, while the area several meters in front of the net and wide is the region of greatest comparative effectiveness for Nadal in the Deuce court. Serves to these regions would generally require explosive movement forward and to the right, suggesting that Nadal and Opelka differ the most in how well they execute this movement when receiving. 

\subsection{How Did Serena Lose the 2019 US Open Women's Final?}

\begin{figure}[h]
    \centering
    \includegraphics[width=0.8\textwidth]{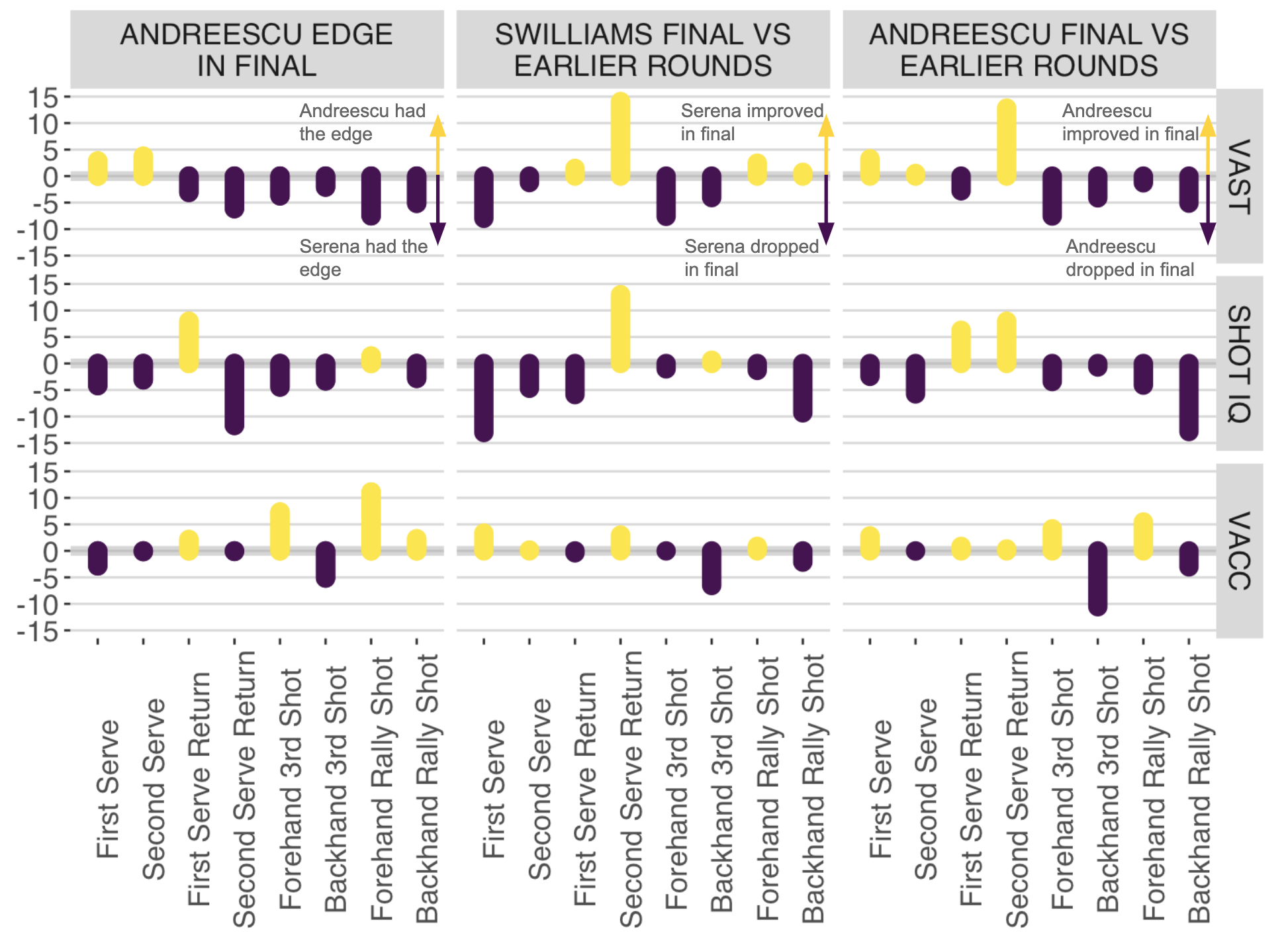}
    \caption{Summaries of the mean VAST, Shot IQ, and VACC for Bianca Andreescu and Serena Williams by major shot types. The left panel shows Andreescu's mean difference vs Serena in the final, the center panel compares Serena's final performance against her average over the previous rounds of the 2019 US Open, the right panel compares Andreescu's final stats against her averages from the previous rounds.}
    \label{fig:womens-final}
\end{figure}

With the groundwork for the ESV metrics now in place, we can return to the match that was our motivating example for the present work: the 2019 US Open women's final. The left panel of Figure~\ref{fig:womens-final} shows the mean differences between Andreescu and Serena on each of our ESV metrics for 8 major shot types. When it came to the difficulty of shots, independent of opponent, the VAST scores show Andreescu had the edge on serve, having an average of +5 percentage points on VAST for first and second serve. Because Andreescu did not have a similar boost on her serve in terms of Shot IQ, we can conclude that the gains in VAST were due more to the strength of execution than shot choice. The area where Andreescu excelled on the return and in rallies was in her court coverage. Against first return and forehand rally shots, in particular, Andreescu was neutralizing shots by several percentage points more than Williams. Thus, while Andreescu didn't have the edge on every metric, we can find areas where she was the better player in each metric category and for each of the most frequent shot types\textemdash first serve, first serve return, and forehand 3rd shot.

Comparing each competitors' performance in the final against their performance in the previous rounds shows how their individual performance may have changed in the last, most critical match of the US Open. For Andreescu (right panel), we can see that she raised her level in terms of her shot taking on first serve and her Shot IQ on the serve return.  For Serena (center panel), some of the most notable changes were her declines in first serve VAST and Shot IQ, each of which were down 10 percentage points or more in the final compared to earlier rounds. The drop in first serve VAST had to be the most crucial for the match, given that the first serve is played in every point and is one of Serena's greatest weapons. In fact, if Serena had dropped just 5 points less in the difficulty of her first serve, she would have been even with Andreescu's first serve difficulty.

\section{Conclusions}

This paper has introduced the VON CRAMM framework for estimating expected shot value in tennis. The key features of the framework are its use of functional encoding of ball and player trajectories, which provide a full-resolution description of the spatiotemporal features of the ball and player movement during a shot that can be well-described with Gaussian mixture models. We have shown that this generative model is the first critical step in a three-step recipe for ESV. Our fully model-based VON CRAMM framework also gives us the machinery to address the problem of attribution in tennis, by enabling us to isolate and quantify the value of key decisions a player makes during shot events. We have demonstrated this by deriving three metrics\textemdash VAST, Shot IQ, and VACC\textemdash that get at the distinct aspects of shot execution, shot choice, and court coverage. Our application of these metrics to the 2019 US Open demonstrates the unique opportunities these advanced metrics can bring for the analysis of high-performance tennis.

While there has been a growing interest in machine learning approaches to address the complexity of tracking data in sport, model-based approaches with functional data have received less attention. We think this has been a missed opportunity for sports analytics given the strong inferential and computational advantages of model-based methods. We hope the introduction of VON CRAMM will encourage the development of similar functional approaches in other sports. We also appreciate that the spatiotemporal complexity of actions in some sports may not be well-described by standard models. For this reason, the combination of high-resolution functional encoding of actions with deep generative models would be an important and novel direction of research for sports analytics. This combination of tools could also be the most flexible in dealing with the next generation of tracking and sensor data and help the analytics community quickly translate new data types into actionable analysis. 

\section*{Acknowledgements}

The authors would like to thank Tennis Australia for providing the tracking data that made this research possible.

\bibliographystyle{plainnat}

\end{document}